\documentclass{aastex}
\usepackage{spr-astr-addons}

\def\bea{\begin{eqnarray}}
\def\eea{\end{eqnarray}}
\def\be{\begin{equation}}
\def\ee{\end{equation}}
\def\nn{\nonumber}
\def\p{\partial}
\def\rr{{\quad r\to r_+\over} \!\!\!\rightarrow}

\begin{document}

\title{Hawking radiation from rotating AdS black holes in conformal gravity}

\author{Shuang-Qing Wu, Gao-Ming Deng\altaffilmark{1} and Di Wu}
\altaffiltext{1}{Institute of Theoretical Physics, China West Normal University, Nanchong 637002, Sichuan, China}
\affil{e-mail: gmd2014cwnu@126.com}

\begin{abstract}
We extend to study Hawking radiation via tunneling in conformal gravity. We adopt Parikh-Wilczek's semi-classical tunneling
method and the method of complex-path integral to investigate Hawking radiation from new rotating AdS black holes in conformal
gravity. In this paper, the research on Hawking radiation from the rotating black holes is done in a general system, not
limited in dragging coordinate systems any longer. Moreover, there existed some shortcomings in the previous derivation of
geodesic equations. Different from the massless case, they used a different approach to derive the geodesic equation of the
massive particles. Even the treatment was inconsistent with the variation principle of action. To remedy the shortcoming, we
improve treatment to deduce the geodesic equations of massive and massless particles in a unified and self-consistent way.
In addition, we also recover the Hawking temperature resorting to the complex-path integral method.
\end{abstract}

\keywords{Hawking radiation, Conformal gravity, Geodesic equation, Tunneling probability}

\section{Introduction}

Conformal gravity, which is an invariant gravity theory under the Weyl conformal transformation, had been advanced
\citep{Englert:1976ep} for a long time. As an intriguing higher-derivative theory \citep{Shin:1989zh,Flanagan:2006ra}, it was
initially hoped for an alternative candidate for a renormalizable quantum gravity beyond Einstein's theory of general relativity,
and had attracted a lot of attention \citep{Jackiw:1978ar,Mansouri:1978pr,Binegar:1982ku,Riegert:1984zz,Fradkin:1984vj,
Horne:1988jf,Novozhilov:1988sn,Mannheim:1988dj,Elizondo:1994vh} during the past decades. An attractive property of conformal
gravity is that it  is a gauge covariant theory \citep{Jackiw:1978ar,Mansouri:1978pr,Horne:1988jf} which shares many similar
characters to that of Yang-Mills gauge theory. In particular, with exact static black hole solutions to conformal gravity found
in \citep{Riegert:1984zz,Mannheim:1988dj}, Birkhoff's theorem was investigated \citep{Riegert:1984zz} and an explanation of
galactic rotating curves was made \citep{Mannheim:1988dj} in the context of the theory. In recent years, renewed interest in
conformal gravity theory has stirred people's nerves once again \citep{Maldacena:2011mk,Fabbri:2011zz,O'Brien:2011wg,Nesbet:2012CGD,
Mannheim:2012qw,Sultana:2012qp,Lu:2012xu,Dunajski:2013zta,Cattani:2013dla}. Among these investigations, there is an attempt to
give an interpretation of the accelerating expansion of the Universe without introducing dark matter and dark energy
\citep{Nesbet:2012CGD}. Meanwhile, with the help of exact vacuum static solution to conformal gravity, the rotating curves of
galaxies \citep{O'Brien:2011wg,Mannheim:2012qw} and the perihelion precession \citep{Sultana:2012qp} as well as light deflection
\citep{Cattani:2013dla} have been researched once more.

It is established that any (A)dS black hole solution in pure Einstein gravity can be embedded into that of conformal gravity.
Although static black hole solution to conformal gravity was known \citep{Riegert:1984zz,Mannheim:1988dj} long ago, the
construction of a rotating AdS black hole solution remains obscure until recently. Therefore, it is of particular interest to
obtain new rotating black holes in conformal gravity that is beyond Einstein's gravity theory. Only quite recently, an exact
solution that describes a rotating and charged AdS black hole has been constructed in \citep{Liu:2012xn} where its global
structure and some thermodynamical properties with an unusual Bekenstein-Smarr formula are analyzed.

As an intriguing viewpoint, regarding the Hawking radiation \citep{Hawking:1974sw} as a tunneling process initiated by Kraus
and Wilczk \citep{Kraus:1994by,Kraus:1994fj} has received so popular attentions that it can be applicable in various black holes.
However, seen from large numbers of previous works concentrating on Hawking radiation as tunneling, they were almost limited in
Einstein gravity, Horava-Lifshitz gravity and the like, unfortunately, research on the tunneling radiation in conformal gravity
was a glaring blank so far. Instead, understanding about Hawking radiation would be more perfect if only extending to study that
in conformal gravity. What's more, as for the new rotating AdS black hole \citep{Liu:2012xn}, some other properties, for example
quantum thermal effect etc., are so attractive that deserving further study. Motivated by these facts, in this paper one of focuses
is on extending to explore the Hawking radiation as tunneling from the new rotating AdS black hole in conformal gravity, thus
filling the blank.

As far as the relatively recent research on the tunneling picture of Hawking radiation is concerned, two different approaches are
currently very popular, one is the Parikh-Wilczek's semi-classical tunneling method \citep{Parikh:1999mf} (based upon the early
developments \citep{Kraus:1994by,Kraus:1994fj,KeskiVakkuri:1996xp}), the other is the Hamilton-Jacobi tunneling method
\citep{Angheben:2005rm} which is developed from the method of complex-path integral proposed in \citep{Srinivasan:1998ty,Shankaranarayanan:2000qv}.
Soon after the appearance of Parikh-Wilczek's work \citep{Parikh:1999mf} on the semi-classical treatment of Hawking radiation as
tunneling, a large amount of work has been focused on directly applying it to various cases of black holes such as those in de Sitter
\citep{Parikh:2002qh,Medved:2002zj,Zhang:2005sf,Huaifan:2009nf}, anti-de Sitter \citep{Hemming:2000as,Liu:2005hj,Wu:2006pz} space-times,
static \citep{Vagenas:2001sm,Vagenas:2002hs,Zhang:2005xt,Jiang:2005xb,Jiang:2006ea,Sarkar:2007sx,Matsuno:2011ca,Kim:2011fh} and
spherical symmetric \citep{Chakraborty:2010di}, rotating charged \citep{Jiang:2005ba,Ali:2007sh,Umetsu:2010kw,LiHui-Ling}, accelerating
and rotating \citep{Rehman:2010zs,Gillani:2011dj} black holes in Einstein-Maxwell theory, black holes in Einstein-Maxwell-dilaton
gravity \citep{Slavov:2012mv}, Black Holes in Horava-Lifshitz gravity \citep{Chen:2009bja,Peng:2009uh}, and those in Einstein-Gauss-Bonnet
gravity \citep{Muneyuki:2011jm} as well as dynamic black holes in noncommutative gravity \citep{Mehdipour:2010ap,Nozari:2012bq}. Although
both methods are used to calculate the tunneling probability of Hawking radiation, they are essentially different in one important aspect:
the Parikh-Wilczek's semi-classical tunneling method considers the fluctuation of the background spacetime with the total conserved
quantities (mass, charge, and angular momentum) fixed, while the method of complex-path integral neglects the back reaction on the black
hole geometry.

Parikh-Wilczek's semi-classical tunneling method is a popular treatment for studying Hawking radiation via tunneling from rotating black
holes, with its help the gravitation effects need not to be taken into account in our discussion. However, an important issue deserves our
more attentions, namely, the frame-dragging effect of a rotating black hole. In general, as a result of a frame-dragging effect of the
coordinate system in the rotating spacetime, the matter field in the ergosphere near the horizon must be dragged by the gravitational field
with an azimuthal angular velocity, so a legitimate physical picture should be described in the dragging coordinate system. Moreover, due to
the presence of rotation, the event horizon does not coincide with the infinite red-shift surface, the geometrical optical limit cannot be
used there since the Kraus-Parikh-Wilczek's analysis is essentially akin to a WKB (`$s$-wave') approximation. Fortunately, this superficial
difficulty can be easily overcome by performing a dragging coordinate transformation which makes the event horizon coincide with the infinite
red-shift surface so that the WKB approximation can be applied now. Therefore, Parikh-Wilczek's semi-classical tunneling method will be used
in the dragging coordinate system to investigate the tunneling process of the rotating black hole in conformal gravity \citep{Liu:2012xn} in
this paper.

With regard to the method of complex-path integral, it is only focused on the leading term to calculate the tunneling probability of Hawking
radiation while the background spacetime is considered to be fixed to simplify the integration of the Hamilton's principal function. However,
in the method of complex-path integral \citep{Angheben:2005rm,DiCriscienzo:2010vz,Rahman:2012id}, one often employs a dragging coordinate
system rather than a non-dragging one to study Hawking radiation from rotating black holes. However it is reasonable that investigations on
the Hawking radiation could have still been completed in a generic non-dragging coordinate system without using a dragging one. Motivated
by this, as an improvement, our one attempt is focused on using it without a coordinate transformation, even for rotating black holes. Actually,
it's available. We will apply this method in a non-dragging coordinate system to calculating the tunneling probability of rotating AdS black holes
in conformal gravity. For the sake of contrast, in the present paper we will investigate the tunneling radiation by means of the Parikh-Wilczek's
semi-classical tunneling method in a dragging coordinate system and the method of complex-path integral without a dragging coordinate transformation
in turn. What's more, it is worth noting that the trick of applying the first law of black hole thermodynamics will greatly simplify the tunneling
integration calculation.

By the way, to facilitate investigation of the tunneling radiation from rotating AdS black holes in conformal gravity, we will introduce a superior
coordinate transformation from various coordinate transformations and recast the new neutral black hole solution into a beautiful form which can
apparently satisfy Landau's condition of the coordinate clock synchronization \citep{CTOF} in the dragging coordinate system, so that the simultaneity
of coordinate clocks could be transmitted from one place to another and has nothing to do with the integration path. In quantum mechanics, particle
tunneling across a barrier is an instantaneous process, therefore this property is necessary for us to investigate the tunneling process. In contrast
to this, in the Appendix we will show another kind of the coordinate transformation which can not fulfill Landau's condition of the clock synchronization.

The derivation of particles' geodesic equations is an important aspect in studying Hawking radiation. However, there existed some shortcomings in previous
derivation. For the case of massive particles, the geodesics were derived by adopting the relation $v_p = v_g/2$ between the group velocity and the phase
velocity, which is inconsistent with the variation principle of action. According to the variation principle of action, in General Relativity, geodesic
equations should be defined by applying the variation principle on the Lagrangian action. What's more, the above treatment is not applicable for defining
the geodesics of massless particles. Differently, the geodesic equation of the massless particles used to be derived by using $d{s^2} = 0$. In this paper,
we will first get the geodesic equation of massive particles by applying the variation principle on the Lagrangian action. Then the massless particles'
geodesic equation can be derived by taking an proper limit of the massive case. Therefore, the geodesic equations of massive and massless particles are
derived in a unified and self-consistent way \citep{WuDi13}.

Our paper is organized as follows. To begin with, we shall review the rotating neutral solution and its thermodynamics of AdS black holes \citep{Liu:2012xn}
recently constructed in the context of conformal gravity in Section \ref{RAbh}. After that, geodesic equations of massive and massless particles will be
derived in a unified and self-consistent way in Section \ref{gomam}. As an example, in Section \ref{Tpop} we will calculate the tunneling probability of
the rotating AdS black holes by adopting the Parikh-Wilczek's semi-classical tunneling method in a dragging coordinate system and the method of complex-path
integral in a non-dragging one respectively. In Section \ref{Htat}, we shall work out the Hawking temperature by means of complex-path integral and make a
comparison with the temperature given in \citep{Liu:2012xn} via the standard method. Section \ref{Conclusion} is devoted to our conclusions and discussion.
The Appendix will give another kind of coordinate transformation.
\section{Rotating AdS black holes in conformal gravity and its thermodynamics\label{RAbh}}

Recently, a new rotating charged AdS black hole solution in conformal gravity that is beyond Einstein gravity was constructed in \citep{Liu:2012xn}. In
this section, we will review the thermodynamical aspect of the neutral rotating charged AdS black hole that is relevant to our research. Introducing
the coordinate transformations $t \rightarrow \tilde{t}$, $\phi \rightarrow \tilde{\phi} -ag^2\tilde{t}$, the neutral rotating AdS black hole solution
can be written as
\bea
d\tilde{s}^2 \hspace{-0.3cm}&= -\frac{\Delta_r}{\Sigma\Xi^2}(\Delta_\theta d\tilde{t} -a\sin^2\theta d\tilde{\phi})^2
 +\Sigma\Big(\frac{dr^2}{\Delta_r} +\frac{d\theta^2}{\Delta_\theta}\Big) \nn \\
& +\frac{\Delta_\theta \sin^2\theta}{\Sigma\Xi^2}\big[a(1 +g^2r^2)d\tilde{t} -(r^2 +a^2)d\tilde{\phi}\big]^2, \qquad
\label{original metric3}
\eea
where
\bea
\hspace{-0.2cm}&& \Sigma = r^2 +a^2\cos^2\theta \, , \quad \Delta_{\theta} = 1 -g^2a^2\cos^2\theta \, , \nn \\
\hspace{-0.2cm}&& \Delta_r = (r^2 +a^2)(1 +g^2r^2) -2\mu r^3 \, , \quad \Xi = 1 -g^2a^2 \, , \nn
\eea
in which $a$, $\mu$ and $g^2$ are parameters related to the mass, angular momentum and the cosmological constant $\Lambda = -3g^2$ respectively.

Accompanying with this new neutral rotating black hole solution, thermodynamical quantities were also given in \citep{Liu:2012xn}. The entropy
and temperature are given by
\bea
&& T = \frac{r_+^2(g^2r_+^2 -1) -a^2(3 +g^2r_+^2)}{4\pi r_+(r_+^2 +a^2)} \, , \label{Htsstandard} \\
&& S = -\frac{2\pi\alpha a^2(1 +g^2r_+^2)}{\Xi r_+^2}\, .
\label{Htsstandard1}
\eea
Note that the parameter $\alpha$ should be negative for a positive entropy in this neutral black hole. The energy and the angular momentum are
computed as
\be
E = -\frac{2\alpha a^2g^2\mu}{\Xi^2} \, , \quad J = -\frac{2\alpha a\mu}{\Xi^2} \, ,
\ee
and the angular velocity at the horizon is
\be
\Omega = \frac{a(1 +g^2r_+^2)}{(r_+^2 +a^2)} \, .¡¡
\ee
Conjugate to the cosmological constant $\Lambda$, the corresponding thermodynamical potential $\Theta$ is given by
\be
\Theta = \frac{\alpha a^2(r_+^2 +a^2)(1 +g^2r_+^2)}{6\Xi^2r_+^3} \, .
\ee
These quantities obey the differential form of the first law of black hole thermodynamics
\be
dE = TdS +\Omega dJ +\Theta d\Lambda \, ,
\ee
while the so-called ``Smarr formula" in Ref. \citep{Liu:2012xn} takes an unusual form
\be
E = 2\Theta\Lambda \, ,
\ee
where $\Lambda$ is a thermodynamic variable resembles the case in paper \citep{Dolan:2013ft}. It can be verified that this formula is indeed valid,
whilst the usual integral Smarr formula can not hold true anymore in conformal gravity theory. This feature might hint an important difference
\citep{Lu:2012xu} of conformal gravity from that of the ordinary General Relativity. However, in this paper, we restrict ourself to the case in which
$\Lambda$ is treated as a true constant.

In order to explore Hawking radiation as tunneling from black holes in conformal gravity in a convenient way, it needs to make a proper coordinate
transformation. Undoubtedly, there are various possible coordinate transformations, such as a different one given in the Appendix. Here we choose a
superior one as follows, its superiority will be shown in the dragging coordinate system later on.
\bea
d\tilde{t} =& \hspace{-0.25cm}dt -\frac{\sqrt{r^2 +a^2}\sqrt{(r^2 +a^2)(1 +g^2r^2)-\Delta_r}}{\Delta_r\sqrt{1 +g^2r^2}}dr \, , \qquad \\
d\tilde{\phi} =& \hspace{-0.55cm}d\phi -\frac{a\sqrt{1 +g^2r^2}\sqrt{(r^2 +a^2)(1 +g^2r^2)-\Delta_r}}{\Delta_r\sqrt{r^2 +a^2}}dr \, .
\label{trans1}
\eea
After performing the above coordinate transformations, the line element (\ref{original metric3}) is changed to the following form,
\bea
ds^2 \hspace{-0.35cm} &=& \hspace{-0.35cm}-\frac{(1 +g^2r^2)\Delta_\theta}{\Xi}dt^2 +\frac{\Sigma}{\Delta_\theta}d\theta^2 +\frac{(r^2 +a^2)\sin^2\theta}{\Xi}d\phi^2 \nn \\
&&\hspace{-0.4cm}  +\bigg[\frac{\sqrt{(r^2 +a^2)(1 +g^2r^2) -\Delta_r}}{\sqrt{\Sigma}\Xi}(\Delta_\theta dt -a\sin^2\theta d\phi)\nn \\
&&\hspace{-0.4cm} +\sqrt{\frac{\Sigma}{(r^2 +a^2)(1 +g^2r^2)}}dr\bigg]^2.
\label{transformation-one}
\eea
It is obvious that in the new form (\ref{transformation-one}), the metric is well behaved at the horizon. In addition, if we set $g^2 = 0$ it will
degenerate into a form which is analogous to the Kerr solution \citep{Jiang:2005ba,WuDi13} if expressed in the Painlev\'{e}-Gullstrand coordinate system.
\section{Geodesic equations of tunneling particles \label{gomam}}

In this section, we work out geodesic equations of massive particles in the new neutral rotating AdS black holes in conformal gravity in a non-dragging
coordinate system. Then, we do a similar analysis in the dragging coordinate system.
\subsection{Geodesic equations of tunneling particles in a generic non-dragging coordinate system \label{gnd}}

Associated with the stationary black hole solutions, there, in general, exist three conserved integration constants that can be deduced via the variational
principle. And these will give rise to the null and timelike geodesic equation as well. Here we restrict the Hamiltonian $\mathcal{H} = m g_{\mu\nu}
\dot{x^{\mu}}\dot{x^{\nu}}/2$ to be a constant, namely $\mathcal{H} = -mk/2$, ($k = 0, 1$) in which the constant $k$ can take two different values
corresponding to the 4-velocity normalization condition of null and timelike geodesic, respectively. Adding two more constants $E$ (energy) and $L$
(angular momentum), one can completely determine the geodesic equations of massive particles. In addition, of special interest is that we can derive the
geodesic equations of massless particles if we set $k = 0$, or the rest mass $m = 0$ from the geodesic equations of massive particles. As a consequence,
we can write out geodesic equations of massive and massless particles in a unified way, thus extending the original work \citep{WuDi13} to the case in
conformal gravity.

At first, we plan to derive the geodesic equation of massive particles from the above metric (\ref{transformation-one}) which is expressed in a non-dragging
coordinate system.

According to Chandrasekhar's book \citep{TMTOBH}, the corresponding Lagrangian that we consider for a massive particle is,
\bea
\mathcal{L} \hspace{-0.3cm}&=& \hspace{-0.3cm} \frac{m}{2} g_{\mu\nu}\dot{x^{\mu}}\dot{x^{\nu}} \nn \\
&=& \hspace{-0.3cm}\frac{m}{2} \Bigg\{\hspace{-0.1cm}-\frac{(1 +g^2r^2)\Delta_\theta}{\Xi}\dot{t}^2
 +\frac{\Sigma}{\Delta_\theta}\dot{\theta}^2 +\frac{(r^2 +a^2)\sin^2\theta}{\Xi}\dot{\phi}^2 \nn \\
&& +\bigg[\frac{\sqrt{(r^2 +a^2)(1 +g^2r^2) -\Delta_r}}{\sqrt{\Sigma}\Xi}(\Delta_\theta \dot{t} -a\sin^2\theta \dot{\phi}) \nn \\
&& +\dot{r}\sqrt{\frac{\Sigma}{(r^2 +a^2)(1 +g^2r^2)}}\bigg]^2 \Bigg\} \, . \qquad
\label{LnondragLa}
\eea
in which $m$ is the rest mass of particle. We derive the Lagrangian (\ref{LnondragLa}) from universal theory which is not dependent on specific gravity,
it's also available for conformal gravity. From it the generalized momenta $P_\alpha = \p\mathcal{L}/\p\dot{x}^{\alpha}$ can be deduced and given by
\bea
P_t \hspace{-0.3cm}&=& \hspace{-0.3cm}\frac{m\Delta_\theta}{\Xi} \Bigg\{-(1 +g^2r^2)\dot{t} \nn \\
&& \hspace{-0.4cm}+\dot{r}\sqrt{1 -\frac{\Delta_r}{(r^2 +a^2)(1 +g^2r^2)}} \nn \\
&& \hspace{-0.4cm} +\frac{(r^2 +a^2)(1 +g^2r^2) -\Delta_r}{\Sigma\Xi}(\Delta_\theta\dot{t} -a\sin^2\theta\dot{\phi}) \Bigg\}, \qquad \\
 \label{nondrapt} \nn
P_r \hspace{-0.3cm}&=& \hspace{-0.3cm}m\bigg[\frac{\Sigma}{(r^2 +a^2)(1 +g^2r^2)}\dot{r} \nn \\
&& \hspace{-0.4cm}+\sqrt{1 -\frac{\Delta_r}{(r^2 +a^2)(1 +g^2r^2)}} \frac{\Delta_\theta\dot{t} -a\sin^2\theta\dot{\phi}}{\Xi} \bigg], \quad \\
\label{nondragpr} \nn
\eea
\be
P_\theta = \frac{m\Sigma}{\Delta_\theta}\dot{\theta} \, , \qquad \label{nondragpx}
\ee
\bea
P_\phi \hspace{-0.3cm}&=& \hspace{-0.3cm}\frac{m}{\Xi} \Bigg\{(r^2 +a^2)\sin^2\theta\dot{\phi} \nn \\
&& \hspace{-0.4cm}-a\sin^2\theta \bigg[\dot{r}\sqrt{1 -\frac{\Delta_r}{(r^2 +a^2)(1 +g^2r^2)}}  \nn \\
&& \hspace{-0.4cm}+\frac{(r^2 +a^2)(1 +g^2r^2) -\Delta_r}{\Sigma\Xi}(\Delta_\theta\dot{t} -a\sin^2\theta\dot{\phi})\bigg]\Bigg\}. \nn \\
\label{nondragpy}
\eea
The corresponding Hamiltonian can be obtained via the Legendre transformation,
\be
\mathcal{H} = \dot{t}P_t +\dot{r}P_r +\dot{\theta}P_\theta +\dot{\phi}P_\phi -\mathcal{L}
 = m g_{\mu\nu}\dot{x^{\mu}}\dot{x^{\nu}}/2 \, . \nn
\ee
In consequence, we have
\bea
\mathcal{H} \hspace{-0.3cm}&=& \hspace{-0.3cm}\frac{m}{2} \Bigg\{\hspace{-0.2cm}-\frac{(1 +g^2r^2)\Delta_\theta}{\Xi}\dot{t}^2
 +\frac{\Sigma}{\Delta_\theta}\dot{\theta}^2 +\frac{(r^2 +a^2)\sin^2\theta}{\Xi}\dot{\phi}^2 \nn \\
&& \hspace{-0.4cm}+\bigg[\frac{\sqrt{(r^2 +a^2)(1 +g^2r^2)
 -\Delta_r}}{\sqrt{\Sigma}\Xi}(\Delta_\theta \dot{t} -a\sin^2\theta \dot{\phi}) \nn \\
&& \hspace{-0.4cm}+\dot{r}\sqrt{\frac{\Sigma}{(r^2 +a^2)(1 +g^2r^2)}}\bigg]^2 \Bigg\} \, . \qquad
\label{HnondragLa}
\eea
Here $t$ and $\phi$ act as ignorable coordinates, and they correspond to the conserved generalized momenta $P_t$ and $P_\phi$. Let them be two
integral constants $E$ and $L$ respectively,
\be
P_t = E \, , \quad P_\phi = L \, , \label{nondragc1}
\ee
we also have the 4-velocity normalization condition,
\be
\mathcal{H} = -m k/2 \, . \nn \\
\ee
Solving these three conditions for $\dot{r}$, $\dot{t}$ and $\dot{\phi}$, we can obtain
\bea
\dot{r} \hspace{-0.3cm}&=& \hspace{-0.3cm}\pm\frac{\sqrt{W}}{m\Sigma} \, , \label{ndrdot}\\
\dot{t} \hspace{-0.3cm}&=& \hspace{-0.3cm}\frac{1}{m\Sigma\Delta_r}\Bigg[-(r^2 +a^2)Y +a\Delta_r\Big(L +\frac{aE\sin^2\theta}{\Delta_\theta}\Big) \nn \\
&& \hspace{-0.4cm}\pm \sqrt{\Big(r^2 +a^2 -\frac{\Delta_r}{1 +g^2r^2}\Big)(r^2 +a^2)W}\Bigg] \, , \label{ndtdot} \\
\dot{\phi} \hspace{-0.3cm}&=& \hspace{-0.3cm}\frac{1 +g^2r^2}{m\Sigma\Delta_r}\Bigg\{-a Y
 +\frac{\Delta_r\Delta_\theta\big(L +\frac{aE\sin^2\theta}{\Delta_\theta}\big)}{(1 +g^2r^2)\sin^2\theta} \nn \\
&& \hspace{-0.4cm}\pm a\sqrt{\Big[1 -\frac{\Delta_r}{(r^2 +a^2)(1 +g^2r^2)}\Big]W}\Bigg\} \, , \qquad
\label{ndydot}
\eea
where we denote
\bea
W \hspace{-0.3cm}&=& \hspace{-0.3cm}Y^2 -\frac{\Delta_r(L\Delta_\theta +aE\sin^2\theta)^2}{\Delta_\theta\sin^2\theta} -\Delta_r(m^2k\Sigma +\Delta_\theta P_\theta^2),\qquad \nn \\
Y \hspace{-0.3cm}&=& \hspace{-0.3cm}(r^2 +a^2)E +a L(1 +g^2r^2) \, . \nn
\eea
With these equations in hand, we proceed to derive the geodesic equations of massive particles in a pretty form in a non-dragging coordinate system.
For the radial part, we get
\bea
\bar{r} \hspace{-0.3cm}&=& \hspace{-0.3cm}\frac{dr}{dt} = \frac{\dot{r}}{\dot{t}} \nn \\
&=& \hspace{-0.3cm}\Delta_r\bigg[(r^2 +a^2)\sqrt{1 -\frac{\Delta_r}{(r^2 +a^2)(1 +g^2r^2)}} \nn \\
&& \pm \frac{-(r^2 +a^2)Y +a\Delta_r\big(L +\frac{aE\sin^2\theta}{\Delta_\theta}\big)}{\sqrt{W}}\bigg]^{-1}, \qquad~
\label{massivenondragr}
\eea
and for the angular part,
\bea
\bar{\phi} \hspace{-0.3cm}&=& \hspace{-0.3cm}\frac{d\phi}{dt} = \frac{\dot{\phi}}{\dot{t}} \nn \\
&=& \hspace{-0.3cm}(1 +g^2r^2)\bigg\{-aY +\frac{\Delta_r\Delta_\theta\big(L +\frac{aE\sin^2\theta}{\Delta_\theta}\big)} {(1 +g^2r^2)\sin^2\theta} \nn \\
&& \hspace{-0.3cm}\pm a\sqrt{\Big[1 -\frac{\Delta_r}{(r^2 +a^2)(1 +g^2r^2)}\Big]W}\bigg\}  \nn \\
&&\hspace{-0.4cm} \times \bigg\{-(r^2 +a^2)Y +a\Delta_r\Big(L +\frac{aE\sin^2\theta}{\Delta_\theta}\Big) \nn \\
&&\hspace{-0.4cm} \pm \sqrt{\Big[r^2 +a^2 -\frac{\Delta_r}{1 +g^2r^2}\Big](r^2 +a^2)W}\bigg\}^{-1} \, ,
\label{massivenondragy}
\eea
where the signs $``\pm"$ correspond to the geodesics of ingoing and outgoing particles from the event horizon.

Next, we will also check their asymptotic behavior of the outgoing particles near the event horizon. When $r$ goes to $r_+$, the radial equation
(\ref{massivenondragr}) and angular equation (\ref{massivenondragy}) behave like
\be
\bar{r} = \frac{dr}{dt} \rr \frac{\Delta^{\prime}(r_+)(r -r_+)}{2(r_+^2 +a^2)}
 = \kappa(r -r_+), \label{asymptoticsr1}
\ee
and
\be
\bar{\phi} = \frac{d\phi}{dt} \rr \frac{a(1 +g^2r_+^2)}{r_+^2 +a^2},\label{asymptoticsy1}
\ee
where $\Delta^{\prime}(r_+)$ denotes the first derivative of $\Delta_r$ at horizon radius $r_+$ and $\kappa$ is the surface gravity on the horizon.

\subsection{Geodesic equations of tunneling particles in a dragging coordinate system\label{gd}}

Now we shall calculate geodesic equation of the new neutral rotating AdS black holes in conformal gravity in a dragging coordinate system.

With regard to the metric (\ref{transformation-one}), in order to make the event horizon coincide with the infinite red-shift surface in the dragging
frame, we first perform a coordinate transformation $d\phi = \bar{\Omega} dt$ in which the dragging angular velocity $\bar{\Omega}$ reads
\be
\bar{\Omega}= -\frac{g_{t\phi}}{g_{\phi\phi}}
= \frac{a\Delta_\theta\big[(r^2 +a^2)(1 +g^2r^2) -\Delta_r\big]}{\Delta_\theta(r^2 +a^2)^2
 -\Delta_r a^2\sin^2\theta} \, ,
\ee
then the metric (\ref{transformation-one}) is sent to the following form
\bea
d\hat{s}^2 \hspace{-0.35cm}&=& \hspace{-0.35cm}-\frac{\Delta_r\Delta_\theta\Sigma}{\Delta_\theta(r^2 +a^2)^2 -\Delta_r a^2\sin^2\theta}dt^2 \nn \\
&& \hspace{-0.5cm}+\frac{2\Delta_\theta\Sigma\sqrt{(r^2 +a^2)\big[(r^2 +a^2)(1 +g^2r^2)
 -\Delta_r\big]}dt dr}{\big[\Delta_\theta(r^2 +a^2)^2 -\Delta_r a^2\sin^2\theta\big]\sqrt{1 +g^2r^2}} \nn \\
&& \hspace{-0.5cm}+\frac{\Sigma}{(r^2 +a^2)(1 +g^2r^2)}dr^2 +\frac{\Sigma}{\Delta_\theta}d\theta^2.
\label{draggingtransformation}
\eea
Perfectly, the new form (\ref{draggingtransformation}) satisfies Landau's condition of the coordinate clock synchronization \citep{CTOF}. In other words,
if the simultaneity of coordinate clocks can be transmitted from one place to another and has nothing to do with the integration path, the metric
components in a dragging coordinate system should satisfy \citep{Zhang:2005xt}
\be
\frac{\p}{\p x^j}\Big(-\frac{g_{0i}}{g_{00}}\Big)
 = \frac{\p}{\p x^i}\Big(-\frac{g_{0j}}{g_{00}}\Big) \qquad (i, j = 1, 2, 3) \, .
\label{clockcondition}
\ee
In fact, because the covariant component $g_{t\theta}$ is zero, it only needs to demand the ratio ${g_{tr}}/{g_{tt}}$ be a function which is independent
of the variable $\theta$. Indeed, its consistency turns out to be true.

In addition, in the new form (\ref{draggingtransformation}), there is no any coordinate singularity at the event horizon in the dragging coordinate system,
and what is more, the event horizon and the infinite red-shift surface coincide with each other so that the geometrical optical limit can be applied. These
properties are very advantageous for us to discuss Hawking radiation via tunneling from rotating black holes and work out the tunneling probability at the
event horizon.

Likewise, the corresponding Lagrangian in this dragging system is given by
\bea
\mathcal{L} \hspace{-0.3cm}&=& \hspace{-0.3cm} \frac{m\Sigma}{2}\Bigg\{-\frac{\Delta_r\Delta_\theta}{\Delta_\theta(r^2 +a^2)^2 -\Delta_r a^2\sin^2\theta}\dot{t}^2 \nn \\
&& \hspace{-0.4cm}+2\frac{\Delta_\theta\sqrt{(r^2 +a^2)\big[(r^2 +a^2)(1 +g^2r^2) -\Delta_r\big]}}
{\big[\Delta_\theta(r^2 +a^2)^2 -\Delta_r a^2\sin^2\theta\big]\sqrt{1 +g^2r^2}}\dot{t} \dot{r} \Bigg\} \nn \\
&& \hspace{-0.4cm}+\frac{\dot{r}^2}{(r^2 +a^2)(1 +g^2r^2)} +\frac{\dot{\theta}^2}{\Delta_\theta},
\label{draggingtransformationLa}
\eea
and we can obtain the generalized momenta $P_\alpha = \p\mathcal{L}/\p\dot{x}^{\alpha}$ as follows
\bea
P_t \hspace{-0.3cm}&=& \hspace{-0.3cm}\frac{m\Sigma\Delta_\theta}{\Delta_\theta(r^2 +a^2)^2 -\Delta_r a^2\sin^2\theta} \Bigg\{-\Delta_r\dot{t} \nn \\
&& \hspace{-0.4cm}+\frac{\sqrt{(r^2 +a^2)\big[(r^2 +a^2)(1 +g^2r^2) -\Delta_r\big]}}{\sqrt{1 +g^2r^2}}\dot{r}\Bigg\},\qquad~ \label{dragpt} \\
P_r \hspace{-0.3cm}&=&\hspace{-0.3cm} m\Sigma\Bigg\{\frac{1}{(r^2 +a^2)(1 +g^2r^2)}\dot{r} \nn \\
&& \hspace{-0.4cm}+\frac{\Delta_\theta\sqrt{(r^2 +a^2)\big[(r^2 +a^2)(1 +g^2r^2) -\Delta_r\big]}} {\sqrt{1 +g^2r^2}\big[\Delta_\theta(r^2 +a^2)^2
   -\Delta_r a^2\sin^2\theta\big]}\dot{t}\Bigg\}, \qquad~
\label{dragpr} \\
P_\theta \hspace{-0.3cm}&=& \hspace{-0.3cm}\frac{m\Sigma}{\Delta_\theta}\dot{\theta}. \label{dragpx}
\eea
Hence the Hamiltonian is
\bea
\mathcal{H} \hspace{-0.3cm}&=& \hspace{-0.3cm}\dot{t}P_t +\dot{r}P_r +\dot{\theta}P_\theta +\dot{\phi}P_\phi -\mathcal{L} = m g_{\mu\nu}\dot{x^{\mu}}\dot{x^{\nu}}/2 \nn \\
&=& \hspace{-0.3cm}\frac{m\Sigma}{2}\Bigg\{-\frac{\Delta_r\Delta_\theta}{\Delta_\theta(r^2 +a^2)^2 -\Delta_r a^2\sin^2\theta}\dot{t}^2 \nn \\
&& \hspace{-0.4cm}+2\frac{\Delta_\theta\sqrt{(r^2 +a^2)\big[(r^2 +a^2)(1 +g^2r^2) -\Delta_r\big]}}
   {\big[\Delta_\theta(r^2 +a^2)^2 -\Delta_r a^2\sin^2\theta\big]\sqrt{1 +g^2r^2}}\dot{t} \dot{r} \nn \\
&& \hspace{-0.4cm}+\frac{\dot{r}^2}{(r^2 +a^2)(1 +g^2r^2)} +\frac{\dot{\theta}^2}{\Delta_\theta} \Bigg\}.
\label{dragHamiltonian}
\eea
Similarly, corresponding to the ignorable coordinate $t$ one can introduce a conservation constant $P_t = E$ (energy) and impose the 4-velocity normalization
condition $\mathcal{H} = -mk/2$, ($k = 0, 1$). Solving these equations which corresponds to these two integral constants, we get
\be
\dot{r} = \pm\frac{E\sqrt{\hat{W}}}{m\Sigma}, \label{dragrdot}\\
\ee
\bea
\dot{t} \hspace{-0.15cm}&=& \hspace{-0.3cm}-\frac{E}{m\Delta_r\Delta_\theta\Sigma\Xi}\bigg\{\Delta_\theta(r^2 +a^2)^2 -\Delta_r a^2\sin^2\theta \nn \\
&& \hspace{-0.9cm}\pm\frac{\Delta_\theta\sqrt{(r^2 +a^2)\big[(r^2 +a^2)(1 +g^2r^2) -\Delta_r\big]\hat{W}}}{1 +g^2r^2}\bigg\}, \qquad~ \label{dragtdot}
\eea
in which
\bea
\hat{W} \hspace{-0.3cm}&=& \hspace{-0.3cm}\frac{(r^2 +a^2)(1 +g^2r^2)\big[\Delta_\theta(r^2 +a^2)^2 -\Delta_r a^2\sin^2\theta\big]^2}
   {\Delta_\theta\big[\Delta_\theta(1 +g^2r^2)(r^2 +a^2)^3 -\Delta^2_r a^2\sin^2\theta\big]} \nn \\
&&\hspace{-0.4cm}\times \Big\{1 -\frac{\Delta_r\Delta_\theta(m^2k\Sigma +\Delta_\theta P_\theta^2)} {E^2\big[\Delta_\theta(r^2 +a^2)^2 -\Delta_r a^2\sin^2\theta\big]}\Big\}\, .
\eea
Thus the geodesic equations of massive particles in the dragging coordinate system can be derived from Eqs. (\ref{dragrdot}) and (\ref{dragtdot}),
\bea
\frac{1}{\bar{r}} \hspace{-0.3cm}&=& \hspace{-0.3cm}\frac{dt}{dr} = \frac{\dot{t}}{\dot{r}} \nn \\
&=& \hspace{-0.3cm}\frac{r^2 +a^2}{\Delta_r}\Bigg\{\sqrt{1 -\frac{\Delta_r}{(r^2 +a^2)(1 +g^2r^2)}} \nn \\
&& \hspace{-0.4cm}\pm \frac{\sqrt{1 -\frac{\Delta^2_r a^2\sin^2\theta}{(1 +g^2r^2)(r^2 +a^2)^3\Delta_\theta}}}
 {\sqrt{1 -\frac{\Delta_r\Delta_\theta(m^2k\Sigma +\Delta_\theta P_\theta^2)} {E^2\big[\Delta_\theta(r^2 +a^2)^2 -\Delta_r a^2\sin^2\theta\big]}}} \Bigg\}.
\label{massivedragr1}
\eea
where the signs $``\pm"$ correspond to the geodesic equations of ingoing and outgoing particles from event horizon.

Now we consider the geodesic equation of massless particles in the dragging coordinate system. Let both $m$ and $P_\theta$ approach zero, we find
\bea
\frac{1}{\bar{r}} \hspace{-0.3cm}&=& \hspace{-0.3cm}\frac{dt}{dr} \nn \\
&=& \hspace{-0.3cm}\frac{r^2 +a^2}{\Delta_r}\Bigg\{\sqrt{1 -\frac{\Delta_r}{(r^2 +a^2)(1 +g^2r^2)}} \nn \\
&& \hspace{-0.4cm}\pm \sqrt{1 -\frac{\Delta^2_r a^2\sin^2\theta}{(1 +g^2r^2)(r^2 +a^2)^3\Delta_\theta}} \Bigg\}.
\label{masslessdragr}
\eea
It is worth noting that the result by setting $g^2 = 0$ in (\ref{masslessdragr}) completely coincides with the previous result \citep{Jiang:2005ba}
which is actually the case of massless particles. As a consequence, we can get geodesic equations of massive and massless particles in a unified
way in the conformal gravity.

There is no doubt that the asymptotic behavior of outgoing particles near the event horizon in the dragging coordinate system needs to be studied.
Therefore, on the basis of the radial equation (\ref{massivedragr1}), when $r$ goes to $r_+$, we have
\be
\bar{r} = \frac{dr}{dt} \rr \frac{\Delta^{\prime}(r_+)(r -r_+)}{2(r_+^2 +a^2)}
= \kappa(r -r_+) \, , \label{asymptoticsr2}
\ee
and the angular velocity at the event horizon is given by
\be
\bar{\phi} = \frac{d\phi}{dt} \rr \frac{a(1 +g^2r_+^2)}{r_+^2 +a^2}
= \Omega_+ \, . \label{asymptoticsy2}
\ee

So far, we have in turn worked out the geodesic equations of particles which tunnel across the event horizon of rotating AdS black holes in conformal
gravity in a non-dragging coordinate system and a dragging one. It is worth mentioning that the geodesic equations of particles have been successfully
derived from the Lagrangian (\ref{LnondragLa}), rather than the relation $v_p = v_g/2$ between the group velocity and the phase velocity, especially
for massive particles, we have effectively avoided aforementioned shortcoming in the process of deriving the massive and massless particles' geodesic
equations and deduced geodesic equations of massive and massless particles in a unified and self-consistent way in conformal gravity.

Compare Eqs. (\ref{asymptoticsr2}) and (\ref{asymptoticsy2}) with (\ref{asymptoticsr1}) and (\ref{asymptoticsy1}), it is indicated that the asymptotic
behaviors of tunneling particles across the event horizon in the dragging coordinate system are the same as that in a non-dragging one, that's just what
we would look forward to.
\section{Tunneling probability of particles via adopting different methods and coordinate systems} \label{Tpop}

Parikh-Wilczek's semi-classical tunneling method and the complex-path integral method have been frequently used to explore Hawking radiation ever before.
However, most of works following these two methods did research on Hawking radiation with the help of coordinate transformations such as a dragging coordinate
transformation, especially for the case of rotating black holes \citep{Zhang:2005gja,Zhang:2005wn,Zhang:2005uh,Jiang:2005ba}. For the purpose of comparison,
we will first use the Parikh-Wilczek's semi-classical tunneling method to compute the tunneling rate in a dragging coordinate system. And then, we shall
adopt the complex-path integral method to calculate the tunneling probability without performing a dragging coordinate transformation.
\subsection{Parikh-Wilczek's semi-classical tunneling method in a dragging coordinate system}

At first, we start with the Parikh-Wilczek's semi-classical tunneling method in a dragging coordinate system to derive the tunneling probability. In the
process of Hawking radiation, the energy of black holes would reduce with particles tunneling out to the infinity, meanwhile, the event horizon surface
would shrink. When taking the self-gravitation between particles into consideration, with the permission for mass fluctuation of black holes under the
conditions of energy conservation as well as angular momentum conservation, and assuming that the energy of particles tunneling out black holes is $\omega$,
then the corresponding line element in the dragging coordinate system should be written as
\bea
d\bar{s}^2 \hspace{-0.3cm}&=& \hspace{-0.3cm}-\frac{\bar{\Delta}_r\Delta_\theta\Sigma}{\Delta_\theta(r^2 +a^2)^2 -\bar{\Delta}_r a^2\sin^2\theta}dt^2 \nn \\
&& \hspace{-0.6cm}+\frac{2\Delta_\theta\Sigma\sqrt{(r^2 +a^2)\big[(r^2 +a^2)(1 +g^2r^2) -\bar{\Delta}_r\big]}}
 {\big[\Delta_\theta(r^2 +a^2)^2 -\bar{\Delta}_r a^2\sin^2\theta\big]\sqrt{1 +g^2r^2}}dt dr \nn \\
&& \hspace{-0.6cm}+\frac{\Sigma}{(r^2 +a^2)(1 +g^2r^2)}dr^2 +\frac{\Sigma}{\Delta_\theta}d\theta^2,
\label{tunelingmetric}
\eea
where $\bar{\Delta}_r = (r^2 +a^2)(1 +g^2r^2) +(M -\omega)\Xi^2r^3/(\alpha a^2g^2)$, in which $M$ is the mass of the black hole. In fact, if a particle
carrying an energy $\omega$ tunnel out black hole, the mass and the angular momentum of the rotating AdS black hole in conformal gravity will change to
$(M -\omega)$ and $(M -\omega)/(ag^2)$, respectively.

What's more, the radial geodesic equation of the outgoing particle is given by
\bea
\frac{1}{\bar{r}} \hspace{-0.3cm}&=&\hspace{-0.3cm} \frac{r^2 +a^2}{\bar{\Delta}_r} \Bigg\{\sqrt{1 -\frac{\bar{\Delta}_r}{(r^2 +a^2)(1 +g^2r^2)}} \nn \\
&& \hspace{-0.4cm}+\frac{\sqrt{1 -\frac{{\bar{\Delta}_r}^2 a^2\sin^2\theta}{(1 +g^2r^2)(r^2 +a^2)^3\Delta_\theta}}}
 {\sqrt{1 -\frac{\bar{\Delta}_r\Delta_\theta(m^2k\Sigma +\Delta_\theta P_\theta^2)}{E^2 \big[\Delta_\theta(r^2 +a^2)^2 -\bar{\Delta}_r a^2\sin^2\theta\big]}}} \Bigg\} \, .
\label{massivedragr}
\eea

According to the WKB approximation, the emission rate of the tunneling particle is related to the action of the tunneling process by
\be
\Gamma\sim e^{-2ImS} \, .
\ee
It can be easily seen from the line element (\ref{tunelingmetric}) that $\phi$ is an ignorable coordinate in the Lagrangian function $\mathcal{L}$.
To eliminate this degree of freedom, the imaginary part of the action should be rewritten as
\bea
\textrm{Im} S \hspace{-0.3cm}&=& \hspace{-0.3cm}\textrm{Im} \int_{t_i}^{t_f}\big(\mathcal{L} -P_\phi\dot{\phi}\big)dt \nn \\
&=& \hspace{-0.3cm}\textrm{Im} \Big[\int_{r_i}^{r_f}P_rdr -\int_{\phi_i}^{\phi_f}P_\phi d\phi\Big] \nn \\
&=& \hspace{-0.3cm}\textrm{Im} \Big[\int_{r_i}^{r_f}\int_0^{P_r}dP_r^{\prime}dr -\int_{\phi_i}^{\phi_f}\int_0^{P_\phi}dP_\phi^{\prime}d\phi\Big],
\eea
where, $r_i$ and $r_f$ correspond to the horizon radius before and after the shrinkage, $P_r$ and $P_\phi$ are the canonical momenta conjugate to
$r$ and $\phi$, respectively. To proceed with an explicit calculation, we can utilize the Hamiltonian equations of motion as follows
\bea
\bar{r} \hspace{-0.2cm}&=&\hspace{-0.2cm} \frac{d\mathcal{H}}{dP_r}\Big|_{(r; \phi,P_\phi)} = \frac{d(M -\omega)}{dP_r} \, , \\
\bar{\phi} \hspace{-0.2cm}&=&\hspace{-0.2cm} \frac{d\mathcal{H}}{dP_\phi}\Big|_{(\phi; r,P_r)} = \frac{\Omega}{ag^2}\frac{d(M -\omega)}{dP_\phi} \, ,
\eea
where we have made use of equations $d\mathcal{H}_{(\phi;r,P_r)} = \Omega dJ = \frac{\Omega}{ag^2} d(M -\omega)$ and $P_\phi = J =\frac{M -\omega}{ag^2}$.
On the basis of previous analysis, we have
\bea
\textrm{Im} S \hspace{-0.3cm}&=& \hspace{-0.3cm}\textrm{Im} \int_{r_i}^{r_f}\int_M^{M -\omega}(d\check{\mathcal{H}} -\check{\Omega} dJ)\frac{dr}{\bar{r}} \nn \\
&=& \hspace{-0.3cm}\textrm{Im} \int_{r_i}^{r_f}\int_M^{M -\omega}\Big[d(M -\check{\omega}) -\frac{\check{\Omega}}{ag^2}d(M -\check{\omega})\Big]\frac{dr}{\bar{r}}. \nn \\
\label{tunelingintegral1}
\eea
In order to complete the above integral, we need the help of the asymptotic behavior of the outgoing particles near the event horizon
\be
\bar{r} \approx \check{\kappa}\,(r -\check{r}_+) \, , \qquad \check{\kappa}
= \frac{\bar{\Delta}^{\prime}(\check{r}_+)}{2(\check{r}_+^2 +a^2)} \, ,
\label{asymptoticr}
\ee
where $\check{\kappa}$ denotes the surface gravity on the horizon after the particles emission. The angular velocity is given by
\be
\check{\Omega} = \frac{a(1 +g^2\check{r}_+^2)}{\check{r}_+^2 +a^2} \, ,
\label{asymptotOmega}
\ee
in addition, the entropy is
\be
\check{S} = -\frac{2\pi\alpha a^2(1 +g^2{\check{r}_+}^2)}{\Xi {\check{r}_+}^2} \, .
\label{asymptotEntropy}
\ee
One can verify that these thermodynamical quantities comply with the differential form of the first law of black hole thermodynamics
\citep{Liu:2012xn} when $\Lambda = -3g^2$ is treated as a constant, namely,
\be
d(M -\check{\omega}) = \frac{\check{\kappa}}{2\pi}d\check{S}
+\frac{\check{\Omega}}{ag^2}d(M -\check{\omega}) \, .
\label{thermodynamicfirstlaw}
\ee
Substituting Eqs. (\ref{asymptoticr}) and (\ref{thermodynamicfirstlaw}) into (\ref{tunelingintegral1}), the imaginary part of action can be written as
\bea
\textrm{Im} S \hspace{-0.3cm}&\approx& \hspace{-0.3cm} \textrm{Im}\int_{r_i}^{r_f}\int_M^{M -\omega}\frac{d(M -\check{\omega}) -\frac{\check{\Omega}}{ag^2}d(M -\check{\omega})}{\check{\kappa}(r -\check{r}_+)}
 dr \nn \\
&=& \hspace{-0.3cm}-\frac{1}{2}\int_{S_{BH}(M)}^{S_{BH}(M -\omega)}d\check{S} = -\frac{1}{2}\triangle S_{BH} \, . \label{imaginary1}
\eea
Therefore, the tunneling probability of particles is
\be
\Gamma\sim e^{-2\textrm{Im} S} = e^{\triangle S_{BH}} \, , \label{tunneling1}
\ee
which indicates that the Parikh-Wilczek's result is also true in the conformal gravity.

So far, we have got the tunneling probability of particles via adopting Parikh-Wilczek's semi-classical tunneling method. In contrast to the
conventional treatment frequently used in tunneling integration, the superiority of using the first law of black hole thermodynamics is highlighted.
There is no doubt that just as that in Einstein gravity, the tunneling probability of a rotating AdS black hole in conformal gravity is also related
to the change of Bekenstein-Hawking entropy and the real radiation spectrum from a black hole is not pure thermal any longer. Here our study is done
in a dragging coordinate system, however, it is, in fact, not limited to do the same analysis in such a system, one can proceed the same procedure in
a generic non-dragging system. To be not repeated, we shall not present the details here. Instead, in the following we shall proceed with the
complex-path integral method in a non-dragging coordinate system. It is worth mentioning that our derivation process is different from most of previous
treatments by means of the same method in a dragging coordinate system.
\subsection{Complex-path integral method in a non-dragging coordinate system}

To develop the complex-path integral method, unlike almost all of related works, here much effort is focused on using it to work out the tunneling
probability in a non-dragging coordinate system, thus to avoid a dragging coordinate transformation, even in rotating black hole in conformal gravity.
We now consider a scalar particle moving in the classical black hole with the background spacetime fixed. Therefore, we neglect the particle's
self-gravitation when considering the semi-classical approximation and onl-y the leading contribution. As a consequence, the corresponding action $I$
satisfies the classical Hamilton-Jacobi equation
\be
g^{\mu\nu}\p_\mu I \p_\nu I +m^2 =0 \, . \label{relativisticHJ}
\ee
Substituting the metric components (\ref{original metric3}) into Eq. (\ref{relativisticHJ}), the Hamilton-Jacobi equation becomes
\bea
\hspace{-0.7cm}&&-\frac{\big[(r^2 +a^2)\p_{\tilde{t}} I +a(1 +g^2r^2)\p_{\tilde{\phi}} I\big]^2}{\Delta_r\Sigma} +\frac{\Delta_r}{\Sigma}(\p_r I)^2 \nn \\
\hspace{-0.7cm}&& +\frac{\Delta_\theta}{\Sigma}(\p_\theta I)^2  +\frac{\Big(a\sin\theta\p_{\tilde{t}} I
               +\frac{\Delta_\theta}{\sin\theta}\p_{\tilde{\phi}} I \Big)^2}{\Delta_\theta\Sigma} +m^2 = 0. \qquad
\label{HJ1}
\eea
Use the following ansatz for variable separation
\be
I = -E\tilde{t} +R(r) +X(\theta) +J\tilde{\phi} \, , \label{vaiableseparation}
\ee
accordingly, we denote
\bea
\hspace{-0.7cm}&& \p_{\tilde{t}} I = -E, \quad \p_{\tilde{\phi}} I = J, \quad \p_r I = R^{\prime}(r) = dR(r)/dr, \nn \\
\hspace{-0.7cm}&& \p_\theta I = X^{\prime}(\theta) = dX(\theta)/d\theta. \qquad~~
\label{vaiableseparation1}
\eea
Substituting Eq. (\ref{vaiableseparation1}) into (\ref{HJ1}) and performing separation of variables $r$ and $\theta$, we have, respectively,
for the radial part and the angular part,
\be
\frac{\Big[a(1 +g^2r^2)J -(r^2 +a^2)E\Big]^2}{\Delta_r} - \Delta_r R^{\prime 2}(r)  -m^2r^2 = C, \label{variablesep1} \\
\ee
\be
\Delta_\theta \Big(\frac{J}{\sin\theta} -E\frac{a\sin\theta}{\Delta_\theta}\Big)^2 +\Delta_\theta X^{\prime 2}(\theta)
+m^2a^2\cos^2\theta = C, \label{variablesep2}
\ee
in which $C$ is the separation constant. Solving the radial part for the outgoing particles from Eq. (\ref{variablesep1}), we get
\bea
&& R_+^{\prime}(r) = \nn \\ 
&& \frac{\sqrt{\big[a\big(1 +g^2r^2\big)J -\big(r^2 +a^2\big)E\big]^2 -\Delta_r \big(m^2r^2 +C\big)}}{\Delta_r}. \nn \\
\label{wd1}
\eea
Since the leading contribution near the horizon is dominant, we may use the following near-horizon approximation,
\be
\Delta_r = (r -r_+)\Delta^{\prime}(r_+) +\cdots \, ,
\ee
thus, Eq. (\ref{wd1}) turns into
\be
R_+^{\prime}(r) = -\frac{(r_+^2 +a^2)E -(1 +g^2r_+^2)aJ}{(r -r_+)\Delta^{\prime}(r_+)} \, . \label{wd2}
\ee
Using the asymptotic behavior of the outgoing particles near the event horizon, it is not difficult to get
\be
R_+(r_+) = i\pi\frac{E -\Omega J}{2\kappa} \, .
\ee
Taking into account the contribution of both the ingoing and outgoing particles and noting that $Im I = Im R(r_+)$ \citep{Angheben:2005rm},
the tunneling probability of particles is
\be
\Gamma\sim  e^{-\frac{2\pi}{\kappa}(E -\Omega J)} \, . \label{tunneling2}
\ee
It is obvious that Eq. (\ref{tunneling1}) is different from Eq. (\ref{tunneling2}), the reason for this is that here the particles'
self-gravitation has been neglected when considering the semi-classical approximation and only the leading contribution. This is different
from our previous treatment.

As a conclusion, to calculate the tunneling probability, both the Parikh-Wilczek's semi-classical tunneling method and the complex-path
integral method are available even in conformal gravity. It should be pointed out that attempt to calculate tunneling probability of Hawking
radiation without dragging coordinate transformations turns out to be a success finally. In other words, calculating the tunneling probability
in a non-dragging coordinate system is indeed feasible, not limited in a dragging system any longer.
\section{Hawking temperature by using the method of complex-path integral} \label{Htat}

As a supplementary verification of the feasibility of the complex-path integration method in conformal gravity, in this section we shall use
it to calculate Hawking temperature of rotating AdS black holes in conformal gravity.

For our aim, the neutral rotating AdS black hole solutions in conformal gravity (\ref{original metric3}) is rewritten in a different form
\bea
d\tilde{s}^2 \hspace{-0.3cm}&=&\hspace{-0.3cm} -\frac{\Delta_r\Delta_\theta\Sigma}{(r^2 +a^2)^2\Delta_\theta -\Delta_r
 a^2\sin^2\theta }d\tilde{t}^2 +\frac{\Sigma}{\Delta_r}d r^2 \nn \\
&&\hspace{-0.4cm} +\frac{\Sigma}{\Delta_\theta} d\theta^2 +\frac{[(r^2 +a^2)^2\Delta_\theta -\Delta_r a^2 \sin^2\theta] \sin^2\theta}{\Sigma\Xi^2} \nn \\
&&\hspace{-0.4cm}\times \bigg\{d\tilde{\phi} -\frac{a\Delta_\theta\big[(r^2 +a^2)(1 +g^2r^2) -\Delta_r\big]}
 {(r^2 +a^2)^2\Delta_\theta -\Delta_r a^2 \sin^2\theta} d\tilde{t}\bigg\}^2, \qquad
\label{Htmetric}
\eea
where $\Sigma$, $\Xi$, $\Delta_r$, $\Delta_\theta$ are the same as those given in (\ref{original metric3}). Considering the near-horizon
approximation of the metric, from (\ref{Htmetric}) we have
\bea
d\tilde{s}^2 \hspace{-0.3cm}&=&\hspace{-0.3cm} -\frac{\Sigma_+\Delta^{\prime}(r_+)(r -r_+)}{(r^2 +a^2)^2} d\tilde{t}^2 +\frac{\Sigma_+ ~dr^2}{\Delta^{\prime}(r_+)(r -r_+)} \nn \\
&& \hspace{-0.6cm}+\frac{\Sigma_+}{\Delta_\theta}d\theta^2 +\frac{(r^2 +a^2)^2\Delta_\theta\sin^2\theta}{\Sigma_+\Xi^2}(d\tilde{\phi} -\Omega_+ d\tilde{t})^2, \qquad~
\label{Htasymmetric}
\eea
where $\Sigma_+ = r_+^2 +a^2\cos^2\theta$ and $\Omega_+ = a(1 +g^2r_+^2)/(r_+^2 +a^2)$ is the angular velocity of the horizon. According to
the work \citep{Angheben:2005rm}, we get
\be
\beta = \frac{4\pi(r_+^2 +a^2)}{\Delta^{\prime}(r_+)} = \frac{2\pi}{\kappa} \, ,
\ee
therefore the Hawking temperature is
\be
T = \frac{r_+^2(g^2r_+^2 -1) -a^2(3 +g^2r_+^2)}{4\pi r_+(r_+^2 +a^2)} \, . \label{cpht}
\ee
It agrees with the expression given in (\ref{Htsstandard}) which is derived by means of standard method. To some extent, thus this conclusion
provides further evidence for feasibility of the complex-path integration method in conformal gravity.
\section{Conclusions and Discussion} \label{Conclusion}

In this paper, we have filled the blank and, as a extension, chiefly studied the Hawking radiation as tunneling in conformal gravity, especially
from the new neutral rotating AdS black holes \citep{Liu:2012xn}, by using the Parikh-Wilczek's semi-classical tunneling method and the
complex-path integral method, thus providing a more perfect understanding for Hawking radiation as tunneling.

With rectifying shortcoming of previous works in deducing the geodesic equations, we have derived geodesic equations of massive and massless
particles in a unified and self-consistent way in conformal gravity. In fact, the improved derivation is available in any other gravity theory,
because process of deriving the geodesic equations from the Lagrangian is not dependent on specific gravity theory.

Afterwards, we have adopted two different methods, namely, the Parikh-Wilczek's semi-classical tunneling method and the method of complex-path
integral to work out the tunneling probability, respectively. It should be pointed out that, in this paper, the method of complex-path integral
have been used in a universal coordinate system, not limited in a dragging coordinate system as before any longer even in the rotating case. Our
results show that the tunneling probability when taking into account self-gravitation is related to the change of Bekenstein-Hawking entropy and
the real radiation spectrum from black hole deviates from the pure thermal spectrum, however, the tunneling probability is not directly in
connection with the entropy change as one considers the semi-classical approximation and the leading contribution while neglecting self-gravitation.

It is of interest to extend the present work to probe further into the Hawking radiation from the charged rotating AdS black holes in conformal
gravity. We leave this for the future plan.

\section{Acknowledgements}

This work is supported by the National Natural Science Foundation of China under grant numbers 11275157 and 10975058.

\section*{Appendix}
\addcontentsline{toc}{section}{Appendix}

\renewcommand{\theequation}{A\arabic{equation}}
\setcounter{equation}{0}

In this Appendix, we will show that there is another kind of coordinate system that can do the tunneling calculation.

If we introduce the following coordinate transformations
\bea
\hspace{-0.3cm}&&d\tilde{t} = dt -\frac{\sqrt{r^2 +a^2}\sqrt{(r^2 +a^2)\Delta_\theta -\Delta_r}}
 {\Delta_r\sqrt{\Delta_\theta}}dr \, , \\
\hspace{-0.3cm}&&d\tilde{\phi} = d\phi -\frac{a\sqrt{1 +g^2r^2}\sqrt{(r^2 +a^2)\Delta_\theta -\Delta_r}}
 {\Delta_r\sqrt{(r^2 +a^2)\Delta_\theta}}dr \, ,
\label{transformation4}
\eea
the metric (\ref{original metric3}) turns into a new form
\bea
ds^2 \hspace{-0.35cm}&=&\hspace{-0.35cm} -\Delta_\theta dt^2 +\frac{\Sigma d\theta^2}{\Delta_\theta}
 +\frac{\Delta_\theta (r^2 +a^2) \sin^2\theta}{\Xi^2}(d\phi -a g^2dt)^2 \nn \\
&& \hspace{-0.4cm}+\Big[\frac{\sqrt{\Delta_\theta (r^2 +a^2) -\Delta_r}}{\sqrt{\Sigma}\Xi}(\Delta_\theta dt -a\sin^2\theta d\phi) \nn \\
&& \hspace{-0.4cm}+\sqrt{\frac{\Sigma}{\Delta_\theta (r^2 +a^2)}}dr \Big]^2.
\label{metric4}
\eea
One can repeat the same procedure in the main text to derive the geodesic equations and the tunneling rate of massive particles, which
is omitted here. Now the dragging angular velocity is
\be
\Omega = -\frac{g_{{t}{\phi}}}{g_{{\phi}{\phi}}}
= \frac{a\Delta_\theta\big[(r^2 +a^2)(1 +g^2r^2) -\Delta_r\big]}{\Delta_\theta(r^2 +a^2)^2
 -\Delta_r a^2\sin^2\theta} \, .
\label{anotherdraggingav}
\ee
After performing a dragging coordinate transformation $d\phi = \Omega dt$, the metric (\ref{metric4}) is then changed into the following form
\bea
ds^2 \hspace{-0.3cm}&=&\hspace{-0.3cm} -\frac{\Delta_r\Delta_\theta\Sigma}{\Delta_\theta(r^2 +a^2)^2 -\Delta_r a^2\sin^2\theta}dt^2 \nn \\
&& \hspace{-0.5cm}+\frac{2\Sigma\sqrt{\Delta_\theta(r^2 +a^2)\big[(r^2 +a^2)(1 +g^2r^2) -\Delta_r\big]}}{\Delta_\theta(r^2 +a^2)^2 -\Delta_r a^2\sin^2\theta}dt dr \nn \\
&& \hspace{-0.5cm}+\frac{\Sigma}{(r^2 +a^2)\Delta_\theta}dr^2 +\frac{\Sigma}{\Delta_\theta}d\theta^2. \qquad~
\label{anotherdraggingtransformation}
\eea
The metric (\ref{anotherdraggingtransformation}) components in the dragging coordinate system can't satisfy Eq. (\ref{clockcondition}) due to
\be
\frac{\p}{\p\theta}\Big(-\frac{g_{{t}r}}{g_{{t}{t}}}\Big)
\neq \frac{\p}{\p r}\Big(-\frac{g_{{t}\theta}}{g_{{t}{t}}}\Big) = 0.
\label{clockcondition2}
\ee
Different from the metric (\ref{draggingtransformation}), the metric (\ref{anotherdraggingtransformation}) can't satisfy Landau's condition of
the coordinate clock synchronization \citep{CTOF}, however, one can still finish the tunneling analysis without any difficulty in this coordinate
system too.

\end{document}